# Integrative Model for Interoception and Exteroception: predictive coding, points of modulation, and testable predictions.

**Pranjal Balar**[1] and **Sundeep Kapila**[2]

[1] University College London, United Kingdom; Swasth Research Fellow
[2] Indian Institute of Technology (Bombay), India; Swasth Research Fellow

*Corresponding author: pranjal.balar.24@ucl.ac.uk*

## Abstract

Interoception and exteroception provide continuous feedback about the body and the environment, yet how they are dynamically integrated within a unified predictive coding framework has remained under-specified. This paper develops and empirically validates an integrative predictive coding model that treats interoceptive and exteroceptive inference as parallel hierarchical systems exchanging precision-weighted prediction errors. Within this framework, arbitration between the two streams is governed by relative precision weights (w), computed in ventromedial and orbitofrontal prefrontal (vmPFC, OFC) circuits and integrated within the anterior insula (AIC) and anterior cingulate cortex (ACC). Neuromodulatory and attentional mechanisms, via dopaminergic, noradrenergic, cholinergic, and vagal pathways, dynamically tune these precision parameters, determining momentary dominance of internal versus external cues in shaping affective experience and behavioral output.

Computational simulations of the model reproduced biologically plausible dynamics: prediction errors decayed exponentially while arbitration weights self-normalized toward equilibrium (w ≈ 0.5), demonstrating stable convergence and coherent integration. Simulated anxiety and PTSD profiles, characterized respectively by interoceptive and exteroceptive overweighting, yielded rigid, self-sustaining imbalances (w → 1 or w → 0) and slowed recalibration, mirroring clinical affective rigidity. Empirical application of the arbitration equation to published EEG-fMRI datasets further validated the model, with estimated w values spanning 0.4–0.9 across contexts of visual attention, cardiac coupling, and anxiety, quantitatively mapping observed neurophysiological variations in heartbeat-evoked and sensory-evoked potentials onto precision-weighted arbitration dynamics. The framework contributes a unifying account of how dysregulated precision weighting may underlie anxiety (overweighted interoception) and PTSD (underweighted interoception).

Building on this validation, a proposed experimental paradigm is outlined to test the model's predictions in humans. It examines recalibration across anxiety, neutral, and PTSD groups following targeted interoceptive or exteroceptive therapies, using physiological, behavioral, and neuroimaging markers of precision and arbitration. Key predictions include identifiable neural markers of coherence (vmPFC-OFC-AIC-amygdala dynamics), modulation of heartbeat-evoked potentials by vagal stimulation, and precision-sensitive behavioral signatures in interoceptive-exteroceptive congruency tasks. Together, these findings and proposed study establish a falsifiable, computationally grounded account of embodied affective control, linking precision-weighted inference to both mechanistic and clinical levels of explanation.

# Introduction

The ability to perceive, interpret, and regulate signals from both the body and the environment is foundational to adaptive behavior and affective regulation. Two primary sensory domains underpin this process: interoception (the perception of internal bodily states) and exteroception (the perception of stimuli from the external world). Interoception encompasses signals related to visceral, nociceptive, and homeostatic activity, including pain, temperature, muscular tension, and vasomotor responses (Craig, 2002, 2003). Exteroception, by contrast, captures information from modalities such as vision, touch, and audition, providing the external reference frame within which internal sensations acquire meaning (Craig, 2009; Critchley & Garfinkel, 2017).

These systems function in continual interaction rather than isolation. Interoceptive processing forms the basis for emotional awareness, self-representation, and vulnerability to psychiatric disorders, whereas exteroceptive cues shape how internal sensations are interpreted and regulated within specific contexts (Barrett & Simmons, 2015). Dysregulation in this balance has been implicated in affective disorders: anxiety involves overweighting of interoceptive prediction errors, leading to excessive salience of bodily cues, while depression involves underweighting or blunted interoceptive precision, resulting in attenuated affective experience (Paulus & Stein, 2010; Khalsa et al., 2018). Understanding the mechanisms by which these domains converge is thus essential for explaining the neural and computational bases of emotional adaptation and pathology.

Predictive coding provides a powerful theoretical framework for explaining how the brain integrates internal and external sensory information. It posits that perception arises from hierarchical inference, where top-down predictions are compared against bottom-up sensory inputs, and discrepancies, or prediction errors, drive model updating (Friston, 2010). While this framework has been extended to interoceptive processes (Seth, 2013; Allen & Friston, 2018), most models treat interoception as an independent channel, underemphasizing the role of exteroceptive context in modulating bodily inference. Yet converging evidence indicates that coherent affective experience emerges from the integration of interoceptive and exteroceptive signals within multimodal cortical hubs such as the anterior insula (AIC) and anterior cingulate cortex (ACC) regions that jointly encode salience and precision (Craig, 2009; Critchley & Garfinkel, 2017).

To address this gap, this paper proposes a refined predictive coding framework that explicitly models the interaction between interoceptive and exteroceptive processing streams. Internal and external sensory inputs are conceptualized as parallel hierarchies generating their own prediction errors, which are subsequently arbitrated and integrated according to their precision weights. Emotional experience thus arises from the inferred interoceptive state contextualized by exteroceptive information. This account also delineates potential modulation points, via vagal stimulation, attention training, or neuromodulatory interventions, where precision weighting can be experimentally or therapeutically adjusted.

By situating interoceptive inference within its broader multimodal predictive context, this framework aims to unify current models of embodied cognition and affect while offering empirically testable mechanisms for recalibrating interoceptive-exteroceptive balance in clinical populations.

# Background

### Early conceptualizations of interoception

Modern understanding of interoception derives largely from Craig's (2002, 2003) formulation of the homeostatic afferent system, which framed interoception as the perception of the physiological condition of the body. Craig identified lamina I spinothalamic pathways projecting to the posterior insula as the primary cortical substrate for interoceptive awareness, emphasizing the insula's role in integrating bodily signals into subjective feelings (Craig, 2009). This reconceptualization established interoception as a distinct sensory modality essential for emotional experience and bodily selfhood (Critchley & Garfinkel, 2017; Khalsa et al., 2018). However, these accounts largely focused on ascending signals and did not specify how top-down expectations or exteroceptive context influence interoceptive processing.

**Predictive coding approaches**

The predictive-coding framework (Friston, 2010) introduced a computational account of perception as hierarchical inference: the brain generates predictions and updates them based on mismatches, or prediction errors. Seth (2013) extended this to interoception, coining interoceptive inference, whereby emotions arise from predictions about bodily states and the minimization of associated errors. This formulation provided a mechanistic basis for emotion and regulation, linking predictive precision to adaptive and pathological processes. Allen and Friston (2018) further emphasized precision weighting, the reliability assigned to prediction errors, as dynamically modulated by neuromodulatory systems, accounting for both heightened bodily vigilance and blunted awareness.

**Toward multimodal integration**

Recent evidence demonstrates that interoceptive and exteroceptive signals converge in cortical hubs such as the insula and anterior cingulate cortex, enabling multimodal integration (Craig, 2009; Critchley & Garfinkel, 2017). Exteroceptive cues, including affective touch, auditory, and visual stimuli, modulate interoceptive precision and the salience of bodily prediction errors (Patchitt et al., 2025; Villani et al., 2019; Mirams et al., 2012; Suzuki et al., 2013; Sel, 2014; Chen et al., 2021; Critchley et al., 2004). This convergence supports the view that the brain jointly evaluates internal and external information streams to construct unified affective percepts.

Nevertheless, theoretical models rarely formalize how these streams interact within predictive coding or how precision weighting is neurally implemented. Although contextual modulation has been acknowledged (Seth, 2013; Barrett & Simmons, 2015), explicit mapping of modulation loci, whether through vagus-nerve stimulation, attention training, or cognitive-behavioral strategies, remains incomplete. Consequently, translational applications have lagged behind theoretical advances.

**Clinical implications**

Empirical findings align with precision-weighting hypotheses: anxiety reflects maladaptive overweighting of interoceptive prediction errors, producing hypervigilance to bodily cues (Paulus & Stein, 2010), while depression reflects underweighting and reduced interoceptive–exteroceptive coherence, contributing to emotional blunting (Smith et al., 2020; Barrett et al., 2017; Nord & Garfinkel, 2022; Khalsa et al., 2018). Without integrative frameworks specifying both streams and modulation points, current models remain limited in explanatory and therapeutic scope.

**Summary of limitations in existing models**

1. Isolated focus on interoception with insufficient integration of exteroceptive inputs.
2. Limited clinical mapping; few models identify modulation points within predictive hierarchies.
3. Underdeveloped translational potential and absence of frameworks linking precision dynamics to interventions such as neuromodulation, attention training, or behavioral therapy.

In response, this paper advances a revised predictive-coding framework that models interoceptive and exteroceptive streams in parallel, identifies their convergence in multimodal cortical hubs, and delineates specific sites for modulation, aiming to refine theoretical understanding and guide targeted clinical strategies.

## Evidence for Interoceptive-Exteroceptive Integration

A growing body of evidence demonstrates that internal and external sensory streams are dynamically integrated to construct a coherent sense of bodily self. When interoceptive and exteroceptive cues are temporally aligned, for example, when visual, auditory, or haptic stimuli are synchronized with heartbeat or respiration, participants show enhanced interoceptive accuracy, higher vagal tone, and improved emotion regulation. These findings indicate that congruent external feedback reinforces bodily rhythms and reduces prediction error, supporting perceptual stability and affective regulation (see Appendix A for detailed review).

Conversely, mismatched internal and external signals produce perceptual conflict, forcing compensatory re-weighting between sensory streams. Classical paradigms such as the rubber-hand illusion, false-feedback tasks, and lesion studies reveal that asynchronous or contradictory cues weaken body ownership and alter affective appraisal. Such misalignments are associated with aberrant activity in the anterior insula and cingulate cortex, providing a mechanistic account for anxiety, depression, and dissociative symptoms that arise from maladaptive weighting of bodily versus environmental information (see Appendix A for detailed review).

Neuroanatomically, convergent evidence identifies the insula-ACC network as the primary locus of integration, where interoceptive prediction errors and exteroceptive context cues are compared, weighted, and updated. Attention, arousal, and neuromodulators (notably vagal, dopaminergic, and noradrenergic pathways) modulate this precision weighting, determining which sensory stream dominates under stress or safety. Clinically, interventions that restore interoceptive precision such as vagus-nerve stimulation, breath-synchronized mindfulness, and body-focused therapies rebalance integration and improve emotional coherence (see Appendix A for detailed review).

Dysregulated integration is particularly evident in fear and trauma, where external threat cues acquire excessive precision, driving hypervigilance and intrusive re-experiencing. In PTSD and related conditions, exteroceptive stimuli dominate inference, leading to persistent activation of salience-network regions such as the amygdala, anterior insula, and ACC. Therapeutic and contemplative interventions that enhance interoceptive awareness, such as breath-focused mindfulness or Somatic Experiencing, restore internal weighting by increasing sensitivity to visceral cues and improving tolerance to autonomic fluctuations. By shifting arbitration toward bodily precision and away from externally triggered prediction errors, these approaches promote emotional stability and reduce maladaptive vigilance (see Appendix A for detailed review).

Integration is maintained through an ongoing closed-loop mechanism in which prediction errors that persist after weighting are used to update priors or drive actions that realign sensory inputs with expectations. Within the insula-ACC-vmPFC hierarchy, ascending interoceptive signals and descending predictions continuously adjust one another through dopaminergic and vagal pathways, implementing active inference. This framework explains how both cognitive reappraisal and bodily regulation, such as controlled breathing, minimize prediction error at different hierarchical levels. The flexibility of the salience network enables adaptive transitions between interoceptive and exteroceptive modes of attention; rigidity within this loop manifests as anxious hypervigilance or blunted affect (see Appendix A for detailed review).

Convergent neurophysiological findings support these dynamics. Reciprocal suppression between vmPFC and oPFC reflects alternating dominance of internal versus external inference streams. The anterior insula acts as a mismatch detector, with heightened right AIC activity signaling integration failures, while vmPFC-amygdala coupling indexes adaptive emotional appraisal. Increased interoceptive sensitivity correlates with stronger heartbeat-evoked potentials and insula-ACC connectivity, whereas disrupted coherence patterns are observed in PTSD, anxiety, and depression. Collectively, these neural and physiological markers validate the model's claim that coherent interoceptive-exteroceptive integration underpins emotional stability, and its disruption corresponds to high-noise, dysregulated affective states (see Appendix A for detailed review).

Full empirical detail and quantitative summaries are provided in Appendix A.

# Proposed Integrative Model: A Revised Predictive Coding Framework

The proposed framework formalizes interoceptive and exteroceptive integration within a unified predictive coding architecture. The framework defines how internal (visceroceptive) and external (sensory) information streams interact through precision-weighted prediction errors and how their arbitration shapes perception, emotion, and action selection. processing as parallel predictive hierarchies that exchange precision-weighted prediction errors. Each stream continuously updates beliefs about internal or external states, with arbitration determining their relative influence on perception and emotion.

- Parallel inference: Interoceptive (visceroceptive) signals from posterior insula and thalamus, and exteroceptive (sensory) inputs from primary sensory cortices each generated prediction errors ($\varepsilon$) weighted by precision ($\pi$).

- Arbitration hub: vmPFC/OFC compute comparative precision ($\Pi^{int}$ vs. $\Pi^{ext}$) to determine stream dominance.

- Integration site: AIC/ACC combine weighted signals into unified percepts and regulate emotional salience.

- Neuromodulatory control: Dopamine (precision/confidence), norepinephrine (salience/arousal), acetylcholine (attentional gain), and vagal afferents (autonomic feedback) dynamically tune $\pi$ at the vental tangmental area, locus coerulus, nucleus tractus solitarius, and the basal forebrain.

- Action selection: Behavioral and autonomic outputs minimize expected free energy across both channels, producing adaptive rebalancing.

- Clinical implication: Anxiety = $\pi^{int} > \pi^{ext}$ (interoceptive overweighting). Depression = $\pi^{int} < \pi^{ext}$ (interoceptive underweighting). Interventions (taVNS, mindfulness, CBT) aim to restore $w \approx 0.5$ (balanced arbitration).

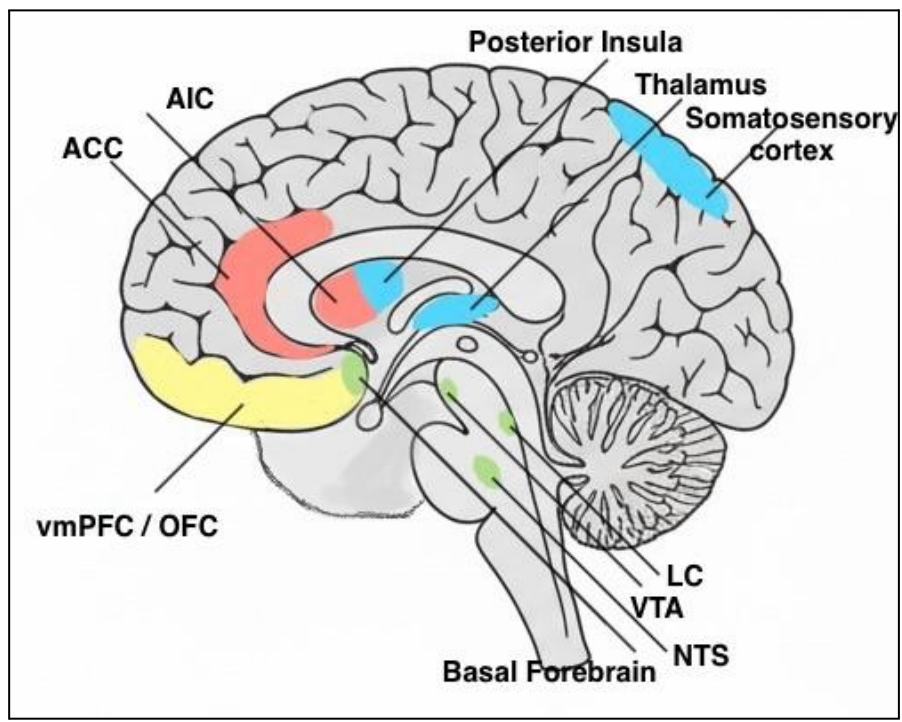

**Figure 1: Brain areas involved in integrating interoceptive and exteroceptive stimuli.**

## Parameter Definitions

| Symbol | Meaning |
|---|---|
| $s^{int}$, $s^{ext}$ | Sensory inputs (interoceptive / exteroceptive) |
| $\hat{\mu}$ | Top-down prediction (prior) |
| $\varepsilon$ | Prediction error (input − prediction) |
| $\pi$ | Precision (inverse variance = confidence weight) |
| $\alpha$ | Learning rate / gain |
| $\lambda$ | Decay constant |
| w | Arbitration weight between internal and external streams |
| a | Action minimizing expected free energy |

## Key Equations

### 1. Prediction errors

$$\varepsilon^{int}(t) = s^{int}(t) - \hat{\mu}^{int}(t)$$

$$\varepsilon^{ext}(t) = s^{ext}(t) - \hat{\mu}^{ext}(t)$$

### 2. Precision-weighted updates

$$\mu^{int}(t+1) = \hat{\mu}^{int}(t) + \alpha^{int} \cdot \pi^{int}(t) \cdot \varepsilon^{int}(t)$$

$$\mu^{ext}(t+1) = \hat{\mu}^{ext}(t) + \alpha^{ext} \cdot \pi^{ext}(t) \cdot \varepsilon^{ext}(t)$$

### 3. Arbitration (weighting by relative precision)

$$\pi^{int\_eff}(t) = \pi^{int}(t) \times \exp(-\lambda \times \varepsilon^{int}(t)^2)$$

$$\pi^{ext\_eff}(t) = \pi^{ext}(t) \times \exp(-\lambda \times \varepsilon^{ext}(t)^2)$$

$$w(t) = \pi^{int\_eff}(t) / (\pi^{int\_eff}(t) + \pi^{ext\_eff}(t))$$

$$\mu^{int-ext}(t) = w(t) \cdot \mu^{int}(t) + (1 - w(t)) \cdot \mu^{ext}(t)$$

### 4. Action selection (active inference form)

$$a^*(t) = \arg\min_a E[(s(a) - \mu^{int-ext})^2]$$

The model's outputs describe how the brain dynamically balances interoceptive-exteroceptive information to form coherent perceptions and guide behavior. Prediction errors ($\varepsilon$) signal mismatches between expected and actual inputs, precision weights ($\pi$) encode confidence in each signal, and the arbitration weight (w) determines which stream, bodily or sensory, dominates perception. Balanced weighting (w ≈ 0.5) supports stable emotion and adaptive regulation, mediated by insula-ACC integration and vmPFC-OFC arbitration, modulated by neuromodulators such as dopamine, norepinephrine, and vagal input. Action ($a^*$) represents the brain's attempt to minimize future prediction errors by acting on the body or environment to make sensations match predictions. When this process fails or becomes biased, maladaptive emotional states emerge (e.g., anxiety, hypervigilance, or emotional blunting).

# Empirical Validation of the Proposed Model

Preliminary validation of the proposed equations was achieved by applying the model to existing neurophysiological datasets that quantify interoceptive and exteroceptive precision. In an EEG-fMRI study of generalized anxiety disorder, markedly elevated heartbeat-evoked potentials (HEP) in the patient group (-2.06 ± 0.53 µV) was reported compared with controls (0.37 ± 0.62 µV; z = 3.92), indicating abnormally amplified interoceptive precision (Verdonk et al., 2023). Substituting these standardized values into the arbitration equation yields $w \approx 0.89$, reflecting a strong interoceptive weighting consistent with hypervigilant bodily focus in anxiety. Conversely, Ren et al. (2024) demonstrated in healthy participants that enhanced cardiac coupling was accompanied by reduced steady-state visual responses (SSVEP phase stability = 0.61 ± 0.13 vs 0.59 ± 0.14; Cohen's d = 0.43) and diminished visual-evoked N2 amplitudes (d = 0.29), while HEP increases were modest (d ≈ 0.43). Applying the same formalism gives $w \approx 0.41$, indicating a relative shift toward exteroceptive precision when visual attention predominates, consistent with reciprocal trade-offs between sensory and visceral streams. Finally, Zaccaro et al. (2024) showed that directing attention to cardiac sensations substantially augmented late HEP positivity (d = 0.89, 352-508 ms; mean = 0.78 across contrasts; the exteroceptive precision value ($\pi_{ex} \approx 0.3$) was approximated as about 40% of the interoceptive magnitude, based on prior HEP attention studies showing markedly weaker exteroceptive modulation during internally focused states), corresponding to $w \approx 0.72$ and confirming interoceptive dominance during internally focused states. Across these studies, empirical w estimates ranged from approximately 0.4 to 0.9, capturing the theoretically predicted continuum from exteroceptive to interoceptive dominance. These findings demonstrate that the model's quantitative formulation accurately maps observed variations in HEP, SSVEP, and VEP dynamics onto the proposed precision-weighted arbitration mechanism, thereby providing preliminary evidence that the equations capture the neural balance underlying coherent interoceptive-exteroceptive integration.

**Table 1: Empirical estimation of interoceptive-exteroceptive arbitration weight (*w*) from reported neurophysiological datasets**

| Reference | Experiment Topic | Reported Values | Computed Parameter(s) | Estimated w |
|---|---|---|---|---|
| Verdonk et al., 2023 | EEG-fMRI study of Generalized Anxiety Disorder (GAD) | Patients: -2.06 ± 0.53 µV; Controls: 0.37 ± 0.62 µV (z = 3.92) | Interoceptive to exteroceptive precision ratio derived from HEP amplitude difference ≈ 5.6 : 0.7 → w = 5.6 / (5.6+0.7) | **0.89** |
| Ren et al., 2024 | Visual-attention modulation of cardiac coupling | SSVEP phase stability: 0.61 ± 0.13 vs 0.59 ± 0.14; HEP: d = 0.43; N2 amplitude: d = 0.29 | Interoceptive to exteroceptive precision estimated from relative HEP-SSVEP magnitudes ≈ 0.43 : 0.61 → w = 0.43 / (0.43 + 0.61) | **0.41** |
| Zaccaro et al., 2024 | Attention to heartbeat sensations | Late HEP positivity (352-508 ms), d = 0.89 (mean ≈ 0.78 across contrasts); baseline exteroceptive precision ≈ 0.3 | Interoceptive to exteroceptive precision ratio derived from mean HEP amplitude ≈ 0.78 : 0.3 → w = 0.78 / (0.78 + 0.3) | **0.72** |

Furthermore, to empirically evaluate the behavior of the proposed interoceptive-exteroceptive arbitration model, a computational simulation was implemented based on the core equations defined in the theoretical framework. Four hierarchical processes were modeled: (1) prediction error computation, (2) precision-weighted updating, (3) precision-based arbitration, and (4) integrated belief formation.

**Simulation setup and assumptions**

The simulation assumed stable learning and noise-free conditions to isolate the intrinsic dynamics of inference and arbitration. Learning rates for both streams were fixed at $\alpha^{int} = \alpha^{ext} = 0.2$, with baseline precisions $\pi^{int} = \pi^{ext} = 0.5$ and a decay parameter $\lambda = 0.2$ governing the exponential down-weighting of large errors. Initial interoceptive and exteroceptive beliefs were set to $\mu^{int}(0) = \mu^{ext}(0) = 0$.

Three environmental conditions were defined to represent different sensory balances: interoceptive dominance ($s^{int} = 0.9$, $s^{ext} = 0.3$), neutral balance ($s^{int} = s^{ext} = 0.5$), and exteroceptive dominance ($s^{int} = 0.3$, $s^{ext} = 0.9$). Each condition was simulated under three starting arbitration weights: $w(0) = 0$ (exteroceptive bias), $w(0) = 0.5$ (balanced), and $w(0) = 1$ (interoceptive bias), yielding nine distinct trajectories. Each simulation ran for 20 iterations, capturing the evolution of prediction errors ($\varepsilon^{int}$, $\varepsilon^{ext}$), arbitration weight ($w$), and integrated belief ($\mu^{int\text{-}ext}$).

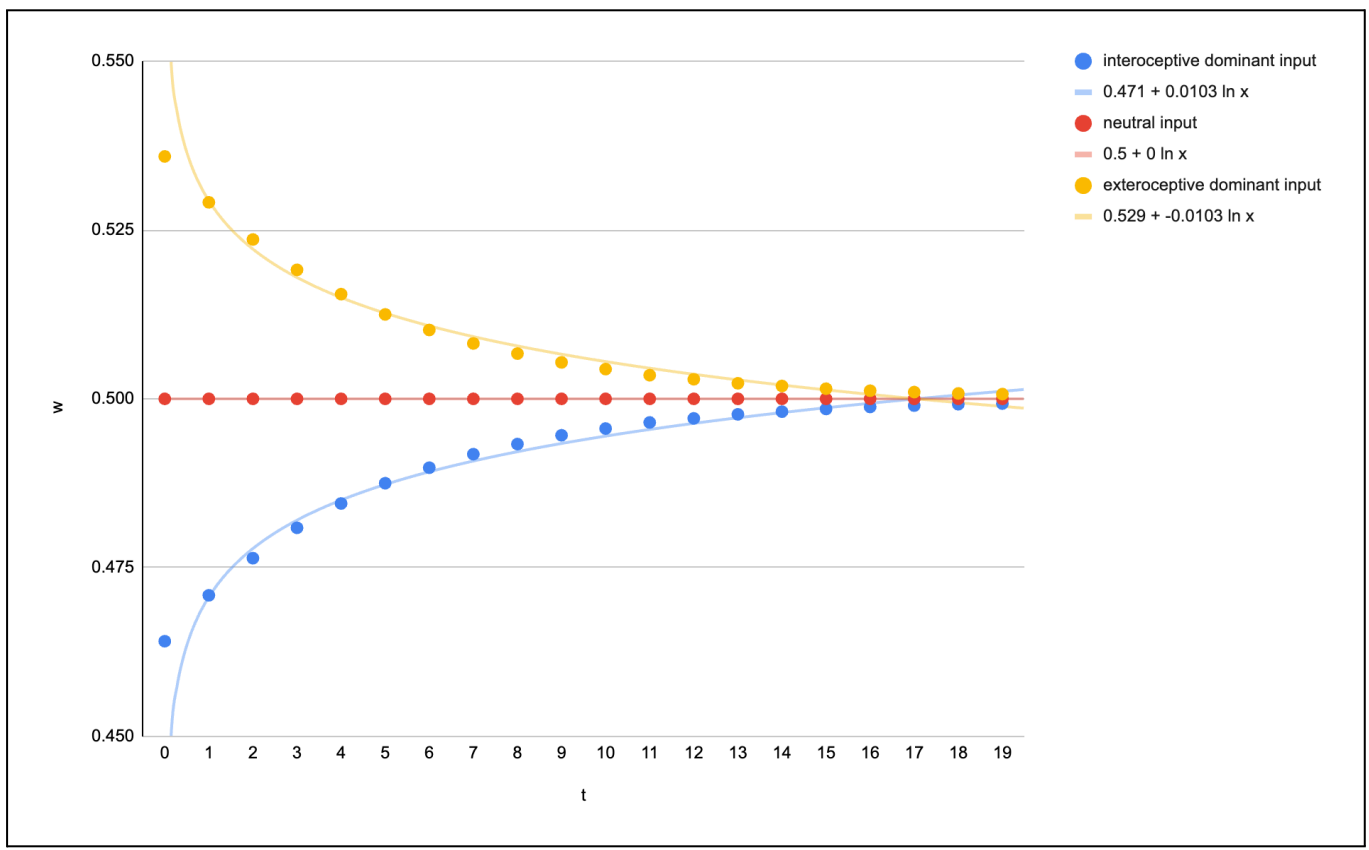

**Figure 2: Dynamic Convergence of Arbitration Weight (w) Across Sensory Conditions** Time-series trajectories of w(t) under interoceptive-dominant, neutral, and exteroceptive-dominant inputs for initial weight w 0. The curves show adaptive reweighting of interoceptive and exteroceptive precision toward a stable equilibrium.

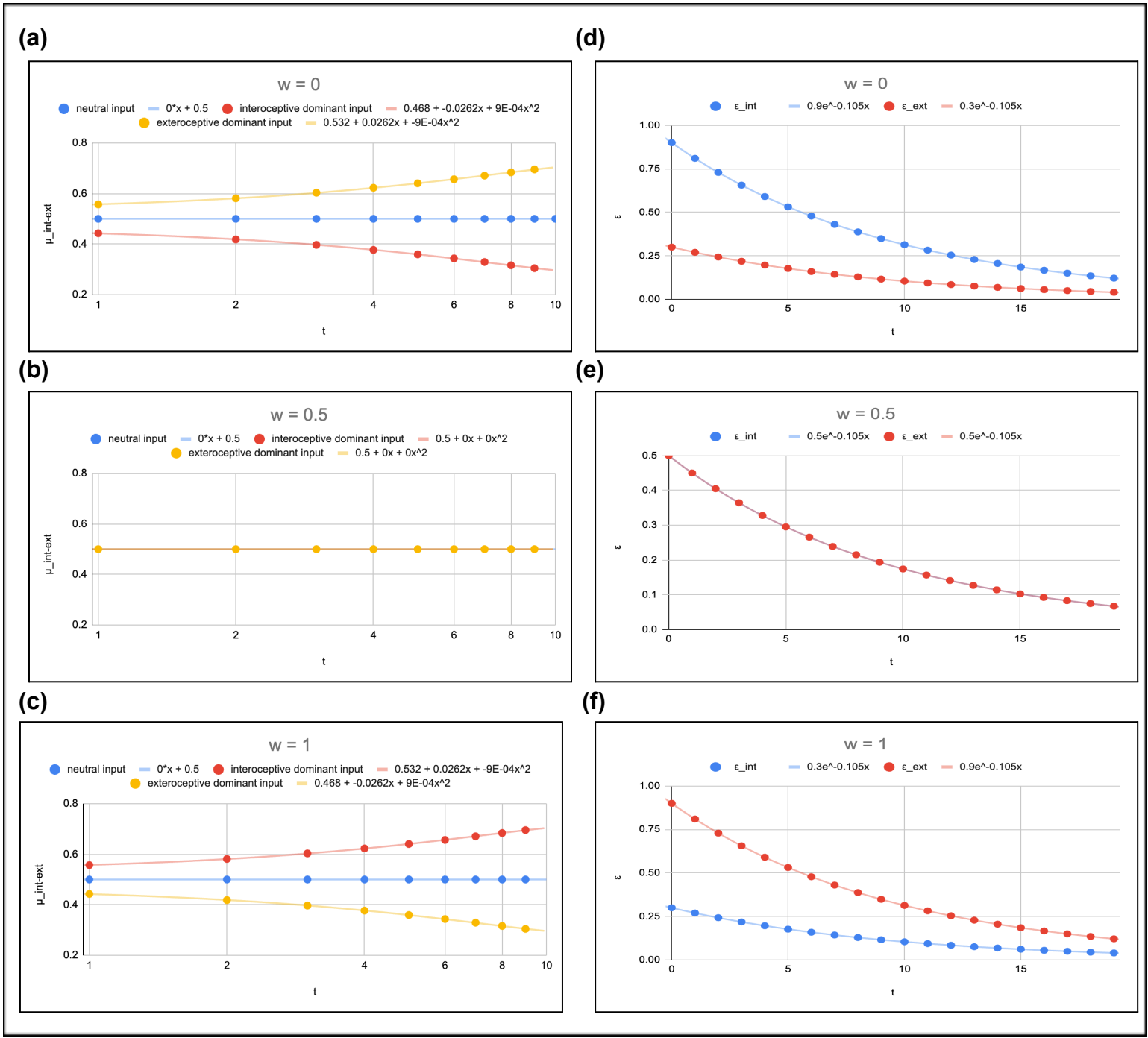

**Figure 3: Integrated belief and prediction-error dynamics under varying arbitration biases** (a-c) Evolution of the integrated belief μ^int-ext(t) across 10 iterations for w = 0, 0.5, 1. (d-f) Corresponding interoceptive (ε^int) and exteroceptive (ε^ext) error trajectories across 20 iterations for w = 0, 0.5, 1. The plots show coupled belief updating and error reduction, with convergence speed shaped by initial weighting and sensory dominance.

**Results and validation**

Across all conditions, the arbitration weight w(t) evolved dynamically toward equilibrium, demonstrating stable convergence consistent with the predictive coding principle of minimizing surprise (Figure 2). The rate and shape of this convergence were modulated by both sensory dominance and initial arbitration bias. In balanced conditions, w(t) stabilized around 0.5, indicating effective precision matching and coherent integration (Figure 2). Under interoceptive-dominant input with an exteroceptive bias ($w_0 = 0$), convergence was slower and nonlinear, with large initial interoceptive errors suppressing effective precision via the $\exp(-\lambda \cdot \varepsilon^2)$ term, producing a "stickier" trajectory (Figure 2). This behavior directly reflects the precision-gating mechanism in the arbitration equation, validating that the exponential term correctly models resistance to updating under large or uncertain prediction errors.

The temporal stabilization of w(t) near 0.5 in most conditions corresponded to the system approaching an integrated belief state, $\mu^{int\text{-}ext}$, as shown in the trajectories of belief updating across initial arbitration biases (Figure 3a-c). These plots illustrate that, as arbitration stabilized, the integrated belief gradually converged toward a coherent perceptual estimate. Concurrently, interoceptive and exteroceptive prediction errors ($\varepsilon^{int}$, $\varepsilon^{ext}$) decreased monotonically across iterations (Figure 3d-f), with the rate of decay varying by sensory dominance and initial weighting. In balanced conditions, $\varepsilon^{int}$ and $\varepsilon^{ext}$ decayed symmetrically, whereas under strong interoceptive or exteroceptive dominance, asymmetrical decay patterns emerged, reflecting precision-dependent learning rates.

## Clinical and psychological interpretation

These dynamics provide a mechanistic account of how interoceptive-exteroceptive balance may underpin affective stability and its disruption in psychopathology. Simulations in which w remains persistently high (w → 1) capture interoceptive overweighting characteristic of anxiety and panic, where bodily signals are afforded excessive precision and dominate conscious appraisal. Conversely, trajectories in which w drifts toward zero (w → 0) reflect PTSD or depersonalization states, where bodily cues are underweighted relative to exteroceptive evidence, leading to blunted emotional awareness and disconnection from bodily feeling.

The nonlinear slowing of rebalancing in extreme initial biases suggests that such states are self-sustaining: large prediction errors reduce effective precision and impede adaptive updating, a phenomenon that may explain clinical rigidity in affective disorders. Interventions such as transcutaneous vagal nerve stimulation (taVNS), mindfulness, or cognitive-behavioral therapy (CBT) can be modeled as parameter changes that increase α or π, or reduce λ, thereby accelerating convergence toward equilibrium (w ≈ 0.5). Empirically, this predicts that therapeutic efficacy should correlate with faster decay of prediction error and restoration of interoceptive-exteroceptive coherence.

## Precision-Weighted Arbitration Dynamics Across Profiles

The dynamic evolution of precision-weighted arbitration (w) between interoceptive and exteroceptive belief streams was investigated across 10 iterative time steps under three distinct sensory dominance conditions - interoceptive-dominant, exteroceptive-dominant, and balanced stimuli, for Neutral, Anxiety, and PTSD profiles (Figure 4). The simulation was designed to assess how systems characterized by altered precision-weighting parameters (such as those observed in anxiety and PTSD) recalibrated or rather failed to recalibrate interoceptive-exteroceptive balance over time.

The model implemented precision-weighted Bayesian updating of internal ($\mu^{int}$) and external ($\mu^{ext}$) beliefs via prediction errors ($\varepsilon^{int}$, $\varepsilon^{ext}$), where each belief stream was adjusted as a function of its precision and learning rate. Effective precisions were dynamically updated through an exponential decay function of squared prediction error, and the arbitration weight w(t) was derived from the relative ratio of interoceptive to total precision. Integrated beliefs ($\mu^{int\text{-}ext}$) were then computed as a weighted combination of the two belief channels.

The initial sensory conditions were defined as follows: interoceptive-dominant ($s^{int}$ = 0.9, $s^{ext}$ = 0.3), exteroceptive-dominant ($s^{int}$ = 0.3, $s^{ext}$ = 0.9), and balanced ($s^{int}$ = $s^{ext}$ = 0.5). For the Neutral profile, learning and decay parameters were set to moderate values (α = 0.3, λ = 0.5, $\pi^{int}$ = $\pi^{ext}$ = 1.0), reflecting adaptive flexibility in sensory precision. The Anxiety profile was characterized by a higher initial interoceptive precision ($\pi^{int}$ > $\pi^{ext}$) and a slower decay constant (λ lower), indicating over-reliance on internal sensory signals. Conversely, the PTSD profile was assigned lower interoceptive precision ($\pi^{int}$ < $\pi^{ext}$) and a faster decay constant (λ higher), producing rapid precision loss and exteroceptive overweighting. Each simulation was run for 10 iterations to observe steady-state convergence or lack thereof.

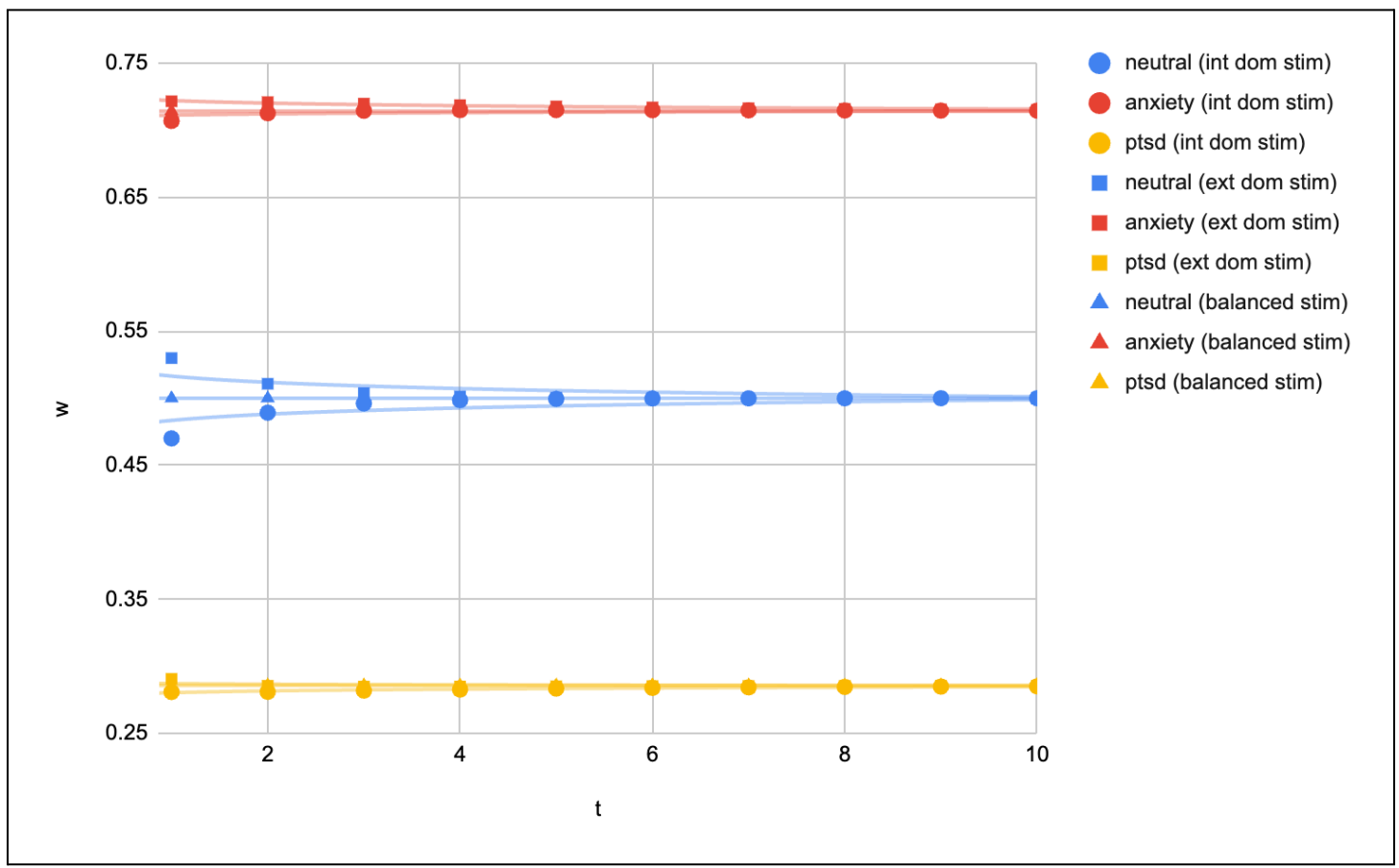

**Figure 4. Temporal evolution of interoceptive weighting (w) under differential sensory dominance across Neutral, Anxiety, and PTSD profiles**

Across all conditions, w(t) reflected the relative weighting of interoceptive precision and served as an index of interoceptive dominance. In the Neutral profile, w values began slightly biased toward the dominant stimulus (≈ 0.47 in interoceptive-dominant and ≈ 0.53 in exteroceptive-dominant conditions) and gradually converged toward 0.5 over successive iterations. This pattern indicated successful adaptive recalibration between interoceptive and exteroceptive inference, reflecting a balanced integration of internal and external sensory information. In contrast, the Anxiety profile exhibited persistently elevated w values (≈ 0.71) across all stimulation types, consistent with excessive interoceptive weighting. The minimal change in w across iterations indicated that this system failed to recalibrate toward equilibrium, maintaining a rigidly interoceptive-dominant state. This rigidity corresponded to maladaptive interoceptive hyperprecision and hypervigilance, characteristic of anxiety phenotypes where bodily sensations are excessively trusted and amplified. The PTSD profile displayed the opposite pattern, with w values remaining consistently low (≈ 0.28) across all conditions. This indicated persistent underweighting of interoceptive information and overreliance on exteroceptive input. Such a pattern corresponded to hypo-interoceptive inference and dissociation from internal states, as commonly observed in post-traumatic stress disorders.

## Clinical Implications and Interventions

The integrative predictive coding framework provides a direct bridge between mechanism and intervention. Because maladaptive emotional states arise from imbalances in interoceptive-exteroceptive precision weighting, effective therapies can be conceptualized as recalibration tools that restore optimal arbitration between internal and external cues (Paulus & Stein, 2010; Khalsa et al., 2018).

**Neuromodulatory interventions**

Transcutaneous auricular vagus nerve stimulation (taVNS) enhances interoceptive precision by strengthening afferent vagal pathways to the nucleus tractus solitarius and insula (Villani et al., 2019; Richter et al., 2021). This modulation increases heartbeat-evoked potentials and normalizes insula-ACC connectivity without altering baseline cardiovascular variables, suggesting a representational rather than purely autonomic effect (Müller et al., 2022). By augmenting precision of interoceptive signals, taVNS supports rebalancing in disorders characterized by excessive exteroceptive weighting, such as PTSD and generalized anxiety.

**Cognitive and behavioural therapies**

Cognitive-behavioural therapy (CBT) and exposure-based approaches can be understood as top-down recalibration strategies that modify maladaptive priors. By systematically confronting distorted predictions about bodily and environmental threat cues, CBT reduces the precision of erroneous exteroceptive priors while strengthening the accuracy of interoceptive appraisal (Barrett & Simmons, 2015).

**Biofeedback and embodied training**

Heart rate variability biofeedback, paced breathing, and mindfulness-based interventions operate as bottom-up entrainment tools that align bodily rhythms with external structure, enhancing coherence and interoceptive trust (Fani et al., 2023; Quadt et al., 2021). These approaches modulate the same salience networks – insula, ACC, and vmPFC-implicated in predictive arbitration, thereby improving emotional stability and bodily awareness.

**Mechanism-to-intervention mapping**

| Mechanistic target | Dysfunction | Intervention strategy | Expected neural outcome |
|---|---|---|---|
| Interoceptive underweighting | Blunted affect, depression | taVNS, slow-breath training | ↑ insula precision, ↑ vagal tone |
| Exteroceptive overweighting | Hypervigilance, anxiety, PTSD | CBT, exposure, mindfulness | ↓ oPFC dominance, ↑ vmPFC-AIC coherence |
| Impaired arbitration | Dissociation, alexithymia | Biofeedback, SE-based titration | ↑ AIC-ACC connectivity, improved HRV |

Interventions that recalibrate precision weighting, either via neuromodulation, cognitive restructuring, or rhythmic entrainment, can restore integrated self-perception and emotional coherence. The framework thus provides a unifying computational rationale for diverse treatment modalities without overextending causal claims.

# Proposed Further Experiments and Empirical Validation

To empirically assess the predictive and translational validity of the proposed interoceptive-exteroceptive arbitration model, an experimental paradigms are proposed.

**Recalibration Across Interoceptive and Exteroceptive Bias Groups**

The experiment aims to test whether groups characterized by distinct interoceptive-exteroceptive biases display differential recalibration trajectories in response to targeted interventions. The study would recruit three populations: individuals with high interoceptive precision bias (clinically anxious or panic-prone), a healthy control group with balanced weighting, and individuals with exteroceptive bias (such as those with post-traumatic stress disorder or depersonalization symptoms). Each group would be assigned to three conditions: interoceptive recalibration, exteroceptive recalibration, or no intervention. Interoceptive recalibration would include protocols such as interoceptive exposure, slow breathing biofeedback, or transcutaneous vagal nerve stimulation (taVNS), each designed to enhance bodily awareness and autonomic regulation. Exteroceptive recalibration would involve sensory grounding or immersive visual-auditory environments intended to increase exteroceptive precision and contextual anchoring.

The key measurable constructs are the model's latent variables of precision ($\pi^{int}$, $\pi^{ext}$), arbitration weight (w), and prediction error ($\varepsilon$). Empirically, these can be approximated through a combination of behavioral, physiological, and neuroimaging measures. Interoceptive precision can be indexed by heartbeat detection accuracy, interoceptive confidence ratings, and heartbeat-evoked potentials recorded via EEG. Exteroceptive precision can be probed using visual or auditory discrimination tasks under uncertainty. Arbitration weight can be estimated from multimodal inference tasks requiring integration of bodily and environmental cues, and quantified via model-based analysis of behavioral weighting and connectivity between anterior insula, anterior cingulate, and ventromedial prefrontal cortex. Prediction error decay can be measured as the rate of reduction in mismatch between expected and observed bodily or sensory states (e.g., HRV reactivity to anticipation).

The model predicts that anxiety-related participants will show high baseline interoceptive precision and slower error decay, resulting in elevated and persistent w values (w > 0.5) (Figure 4). Exteroceptive recalibration should reduce this dominance, normalizing precision ratios and restoring w toward equilibrium. Figure 5 illustrates how a linear increase in the decay constant causes the graph to plateau at w = 0.5, visualising the effect of an intervention on w. In contrast, PTSD-like participants should display low interoceptive precision and underweighting of bodily evidence (w < 0.5) (Figure 4). Interoceptive recalibration should enhance the precision and bring arbitration toward balance. Figure 5 illustrates how a linear decrease in the decay constant causes the graph to plateau at w = 0.5, visualising the effect of an intervention on w. Healthy participants are expected to show stable, symmetric error decay and minimal change. Recalibration techniques are conceptualized as interventions that effectively increase or decrease the decay constant ($\lambda$), thereby enhancing adaptive updating of precision weights and facilitating convergence of w toward 0.5.

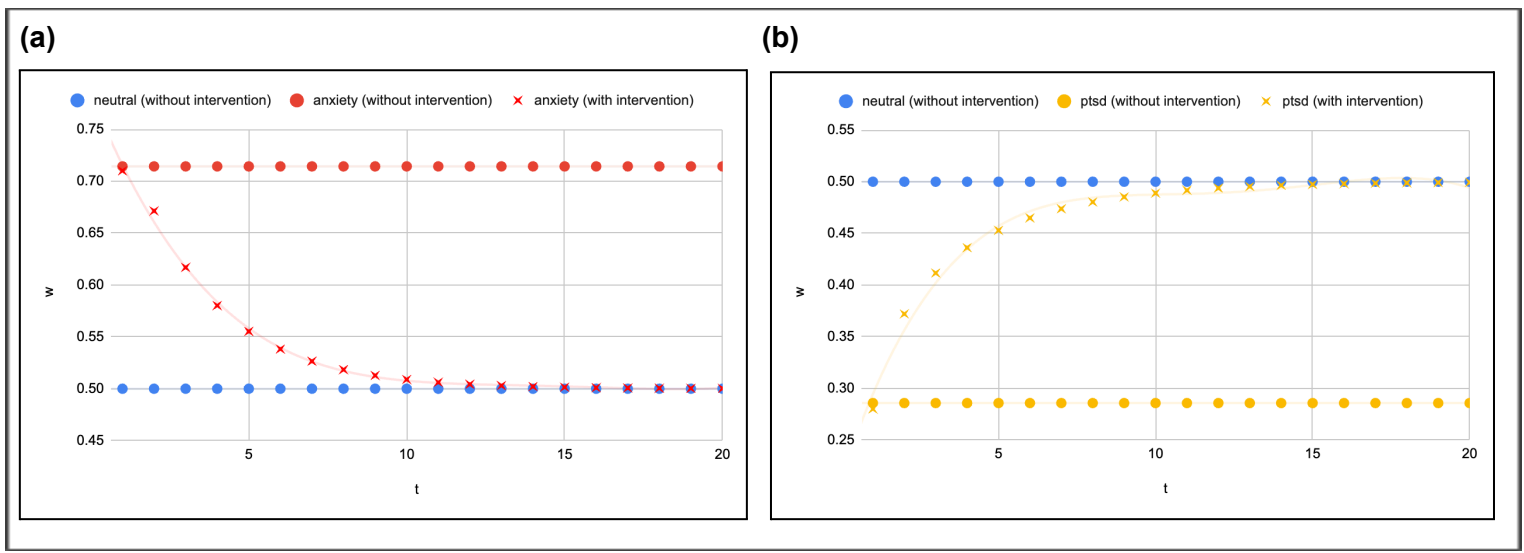

**Figure 5. Temporal evolution of interoceptive weighting (w) for Neutral, Anxiety, and PTSD profiles with or without interventions**

## Conclusions and Future Research

This paper advances a predictive coding framework in which interoceptive and exteroceptive streams operate as parallel but interacting hierarchies. The model formalizes both as parallel hierarchies that exchange prediction errors, with the arbitration weight (w) representing the relative influence of internal and external sensory streams. Simulations showed that prediction errors decrease over time and that w converges toward a stable equilibrium around 0.5 under balanced conditions. Emotional experience and selfhood emerge from the arbitration and integration of these streams, modulated by attention, neuromodulation, and embodied action. When precision is biased toward one stream, the system becomes rigid: high interoceptive weighting (w ≈ 1) represents anxiety-like states, while low interoceptive weighting (w ≈ 0) represents PTSD-like states.

Empirical validation using published EEG and fMRI datasets supported the model. Estimated w values ranged from 0.4 to 0.9 across studies of anxiety, visual attention, and interoceptive focus, as summarized in Table 1. These values matched the theoretical prediction that emotional states vary along a continuum from exteroceptive to interoceptive dominance.

The model also provides a mechanistic view of interventions. Vagal nerve stimulation, mindfulness, and cognitive-behavioral therapy can be described as changes in precision parameters that help restore balanced arbitration (w ≈ 0.5). Simulations suggest that such recalibration increases learning rates, reduces prediction error, and stabilizes integration between internal and external cues.

Future research should prioritise experimental tests of these mechanisms, thus, the paper proposes a follow-up experiment to test these predictions in human participants. The study will compare anxiety, PTSD, and control groups before and after targeted interoceptive or exteroceptive interventions, using behavioral, physiological, and neuroimaging measures to estimate w and track error reduction. This will allow direct testing of whether precision rebalancing through intervention produces measurable changes in arbitration dynamics. Furthermore, developing standardised coherence markers will allow translational application across psychiatry, neurorehabilitation, and embodied cognition research.

# Appendix A

## Evidence for Interoceptive-Exteroceptive Integration

### Aligned integration: coherence and perceptual stability

Aligned integration occurs when interoceptive and exteroceptive streams converge in synchrony, reinforcing one another to produce a coherent bodily self-representation and improved regulation of physiological and emotional states. Empirical evidence shows that synchronizing external cues, whether auditory, visual, or haptic, with internal rhythms such as breathing or heartbeat strengthens the salience of interoceptive signals, enhances predictive accuracy, and stabilizes self-perception.

For instance, vibroacoustically-augmented, breath-focused mindfulness demonstrates how rhythmic exteroceptive stimulation can be used to support interoceptive attention and regulation. By coupling breath-synchronized vibration with mindfulness, participants showed reduced stress, greater heart rate variability (HRV), and neurophysiological patterns associated with relaxation, including reductions in alpha rhythms and increased connectivity within salience-related networks (Fani et al., 2023). These results suggest that when external inputs reinforce bodily rhythms, the brain can recalibrate internal predictions and reduce uncertainty, improving trust in bodily signals and supporting emotion regulation.

Aligned integration is also evident in sensory substitution paradigms, where exteroceptive cues are used to mimic or reinforce interoceptive states. Transcutaneous auricular vagus nerve stimulation (taVNS), for example, enhances interoceptive accuracy on heartbeat discrimination tasks by improving the alignment between cardiac sensations and external auditory tones (Villani et al., 2019). Neural evidence indicates that this effect is mediated by the anterior insula, which integrates internal and external cues while modulating arousal (Villani et al., 2019). False physiological feedback (FFB) experiments extend this principle: when pulsatile somatosensory stimulation is delivered at the wrist to mimic heartbeats, it alters emotional judgments, with faster or slower-than-actual rhythms biasing perceived arousal intensity (Patchitt et al., 2025). Similarly, cardio-visual illusions, where a virtual hand flashes in synchrony with one's heartbeat, strengthen the sense of body ownership, whereas asynchronous feedback weakens it (Suzuki et al., 2013). These findings underscore that synchronous interoceptive-exteroceptive coupling stabilizes bodily self-location and affective evaluation, while misalignment disrupts integration.

Additional evidence shows that low-frequency vibroacoustic stimulation, particularly amplitude-modulated 40 Hz vibration, can alter mood and HRV, suggesting that such rhythmic exteroceptive inputs interact with internal physiological oscillations to reinforce coherence between body and environment (Vilímek et al., 2022). Parallel findings from breathing biofeedback studies highlight the importance of synchrony; slow-paced breathing guided by immersive virtual reality or haptic sensors enhances cardiac coherence and autonomic regulation (Farb et al., 2015; Bouny et al., 2023). By externally structuring respiration at resonance frequency, these interventions reinforce the CNS-ANS loop, improving interoceptive control and emotion regulation through insula and prefrontal engagement.

Aligned integration can also extend into aesthetic and artistic domains. Xu and Cho (2023) demonstrated that visualizing physiological data such as galvanic skin response or photoplethysmography as artistic, real-time feedback increases relaxation and emotional self-awareness. By transforming internal signals into external, perceptible representations, participants became more mindful of their bodily states, supporting prefrontal and visual cortical mechanisms of awareness.

Quantitatively, across these literatures the aligned conditions typically show: (i) significant increases in interoceptive neural markers (e.g., larger HEPs during stimulation/training; stronger task-locked insula responses); (ii) measurable autonomic shifts compatible with higher vagal tone (e.g., increased HRV indices during/after slow-breath and congruent feedback protocols); (iii) behavioral improvements on accuracy/insight tasks (higher correct discrimination, tighter confidence-accuracy coupling); and (iv) symptom reductions of at least small-to-moderate magnitude in anxiety/dissociation scales relative to control practices. Together, these findings suggest:

1. precision weighting is trainable; closed-loop, body-synchronous cues and slow-breath states up-regulate visceral signal precision while reducing exteroceptive noise;
2. arbitration is contextual; AIC/ACC allocate gain more effectively when exteroceptive streams are designed to confirm (not compete with) internal rhythms; and
3. self-environment alignment is implementable; congruent biofeedback and attentional scaffolds act as control inputs that nudge the system toward a single, coherent generative story, reducing prediction error at lower cost to affective stability.

**Unaligned integration: mismatch and compensatory strategies**

Unaligned integration occurs when external (exteroceptive) and internal (interoceptive) bodily signals are incongruent, leading to perceptual conflict and the necessity for compensatory strategies within the brain (Seth et al., 2011; Barrett & Simmons, 2015; Owens et al., 2018).

Experimental paradigms clearly demonstrate this phenomenon. In the rubber-hand illusion (RHI), synchronous affective touch enhances the illusory sense of body ownership, whereas asynchronous stroking reduces it (Crucianelli et al., 2018; Tsakiris et al., 2011). This illustrates how mismatches between exteroceptive (visual and tactile) and interoceptive (affective touch, cardiac awareness) information can alter the sense of bodily self (Suzuki et al., 2013; Park & Blanke, 2019; Moffatt et al., 2024). Similarly, directing attention to interoceptive signals can increase false alarms in tactile detection tasks, suggesting that such mismatches bias perceptual thresholds in the exteroceptive domain (Mirams et al., 2012). This implies that focusing internally might lead to misinterpreting sensory "noise" as a true external signal (Mirams et al., 2012; Brown, 2004).

Neuroimaging and lesion studies provide further evidence for competitive interactions under mismatch conditions. Reciprocal suppression has been observed between interoceptive and exteroceptive cortical regions. Interoceptive (visceroappetitive) feedback preferentially activates the ventromedial prefrontal cortex (VMPFC) while deactivating the orbitofrontal cortex (OPFC), whereas visual exteroceptive feedback reverses this pattern (Hurliman et al., 2005). Amygdala activity correlates closely with the VMPFC during these shifts, reflecting the affective weighting of mismatches (Hurliman et al., 2005). The anterior insula cortex (AIC), a key interoceptive hub, also acts as a comparator, showing heightened responses to false physiological feedback (Gray et al., 2007; Paulus & Stein, 2006). These findings suggest that when sensorimotor integration fails, the brain is compelled to prioritise one sensory stream over the other (Hurliman et al., 2005; Owens et al., 2018), assigning precision weightings based on reliability to maintain optimal perception (Owens et al., 2018; Seth et al., 2013).

Such mismatches and failures in integration provide a mechanistic explanation for symptom patterns in psychiatric conditions (Khalsa et al., 2018; Nord & Garfinkel, 2022; Moffatt et al., 2024). In anxiety, overemphasis on interoceptive prediction error signals can drive catastrophic interpretations of benign exteroceptive contexts (Paulus & Stein, 2006; Dunn et al., 2010; Quadt et al., 2018; Livermore et al., 2024). This heightened focus on internal bodily sensations, often without corresponding improvements in objective accuracy, fuels anxious vigilance (Mirams et al., 2012; Yoris et al., 2015). In contrast, depression is associated with underweighting of interoceptive inputs, leading to muted emotional and physiological responses even in stimulating environments (Barrett, 2017; Rottenberg et al., 2005; Schwerdtfeger & Rosenkaimer, 2011; Harshaw, 2015; Dunne et al., 2021). This is consistent with findings of insula hypoactivation during interoceptive tasks in depressed individuals (Avery et al., 2014; Wiebking et al., 2010; Hu et al., 2023). Dysfunction in interoceptive processing, including discrepancies between interoceptive accuracy and subjective beliefs (Garfinkel et al., 2015), is therefore a common feature across multiple mental health conditions (Khalsa et al., 2018; Nord & Garfinkel, 2022).

**Neural substrates of convergence**

The integration of bodily signals relies on defined neuroanatomical hubs. The insular cortex demonstrates a posterior-to-anterior gradient; the posterior insula encodes granular interoceptive inputs, while the anterior insula (AIC) integrates these signals with top-down predictions and contextual cues (Craig, 2009; Farb et al., 2013; Barrett & Simmons, 2015). Lesion studies highlight its necessity; right insula damage reduces affective touch sensitivity, disrupting the binding of external tactile cues with interoceptive affective quality (Kirsch et al., 2020).

Interoceptive pathways to posterior insula are relatively well described, whereas brainstem relays for exteroceptive inputs targeting pINS are less clearly delineated. Visceral, nociceptive, and thermal signals ascend via lamina I/spinothalamic and vagal/cranial afferents to the nucleus tractus solitarius (NTS). From NTS, projections engage the parabrachial nucleus (PBN) and posterior thalamus (e.g., VMpo/VPI; gustatory via VPMpc) before reaching posterior/granular insula (pINS/gINS), which serves as primary interoceptive cortex and preserves modality-specific maps for homeostatic signals (incl. pain/temperature and taste). (Craig 2002; Craig 2009; Evrard 2019; Fermin et al., 2023). The pINS receives convergent cortico-cortical input from higher-order sensory and association areas, classically SII/somatosensory, auditory belt/parabelt, premotor, and inferior/parietal association cortices, supporting multisensory integration (Kurth et al., 2010; Haruki & Ogawa, 2021; Nieuwenhuys, 2012). A concrete sub-case is vestibular input: the parieto-insular vestibular cortex (PIVC; area OP2) in posterior insula receives thalamic vestibular projections from brainstem vestibular nuclei, providing a partly "direct" exteroceptive route into pINS. Taste within mid/posterior insula (via NTS→PBN→VPMpc) spans interoceptive-exteroceptive boundaries (Faurion et al., 1999). On this basis, pINS/gINS is best framed as an integration hub receiving (i) thalamo-insular interoceptive afferents and (ii) convergent exteroceptive inputs mainly via association cortices (plus vestibular thalamo-insular projections), with AIC linking these representations to salience/valuation circuits (AIC-ACC-OFC/VMPFC) to construct unified feeling states (Craig 2009; Nord & Garfinkel 2022).

Recent models expand on this organisation. The IMAC model (Fermin et al., 2023) proposes that the insula operates as a hierarchical network composed of granular, dysgranular, and agranular modules. This architecture supports a stepwise transformation of interoceptive information, from low-level sensory predictions to higher-order conscious feelings, refining our understanding of how predictive coding mechanisms are anatomically realised.

Convergence extends to the anterior cingulate cortex (ACC), which acts as a salience hub regulating emotional arousal and attentional control (Menon & Uddin, 2010). Functional imaging shows affective touch strengthens connectivity between posterior and anterior insula while engaging the ACC and medial prefrontal cortex (Crucianelli et al., 2018; McGlone et al., 2014). Interventions that shift bodily states, such as vagus nerve stimulation, recalibrate this network by enhancing parasympathetic tone and modulating insula-ACC activity, improving interoceptive accuracy (Villani et al., 2019; Richter et al., 2021).

At the clinical level, Nord & Garfinkel (2022) emphasise how disturbances in these interoceptive-exteroceptive pathways underpin common mental health conditions, and outline how interventions can directly target these neural circuits. This perspective underscores the translational relevance of the insula-ACC network in guiding therapeutic approaches.

Taken together, these findings establish the insula-ACC network as the central cortical substrate where interoceptive and exteroceptive predictions intersect, are weighted, and updated (Seth & Friston, 2016). Dysregulation within this circuitry is therefore a plausible neural basis for altered bodily awareness and affective disturbance in anxiety, depression, and autism (Khalsa et al., 2018; Harshaw, 2015).

**Modulatory factors shaping relative weighting**

**Attention**

Attention dynamically shifts the precision of interoceptive and exteroceptive signals, thereby modulating their integration. Neuroimaging shows that interoceptive attention (IA) engages posterior insula, while exteroceptive attention (EA) engages visual and frontoparietal networks; the anterior insula links both, supporting emotional binding (Farb et al., 2013). Electrophysiology and behaviour confirm this precision tuning: heartbeat-evoked potentials (HEPs) increase under interoceptive focus (Petzschner et al., 2019), and IA biases tactile detection thresholds towards more liberal reporting (Mirams et al., 2012). Within predictive coding, attention operates as a gain control, amplifying the influence of attended prediction errors (Seth, 2013). This stabilises body representations but, when maladaptive, contributes to hypervigilance in anxiety or blunted interoceptive sensitivity in depression (Nord & Garfinkel, 2022).

**Arousal and stress**

Autonomic state and arousal strongly shape how interoceptive and exteroceptive signals are weighted. Bodily states conveyed via interoceptive pathways not only regulate homeostasis but also influence emotion, cognition, and perception (Critchley & Harrison, 2013). Experimental manipulations show that pulsatile somatosensory stimulation mimicking cardiac rhythms alters emotional judgements by modulating interoceptive representations of arousal, engaging the anterior insula (Patchitt et al., 2025). Similarly, false physiological feedback (FFB) at slower-than-actual heart rates reduces anxiety without entraining physiology, indicating that appraisal of combined internal-external signals drives behavioural change (Patchitt et al., 2025). Stress and cardiovascular arousal bias perception toward interoceptive threat signals. Threat detection is selectively enhanced when stimuli are presented at systole, amplifying fearful face perception and fear learning (Garfinkel et al., 2014; Garfinkel et al., 2021). Stress-induced arousal is also linked to suppression of the baroreflex, mediated by insula, ACC, amygdala, and brainstem activity (Gianaros et al., 2012). Interventions targeting vagal pathways can counteract these biases (Villani et al., 2019). Non-invasive vagus nerve stimulation (taVNS) improves cardiac interoceptive accuracy, enhances heartbeat-evoked potentials, and modulates insula-ACC circuitry (Villani et al., 2019; Richter et al., 2021). Broader interoceptive training and mindfulness-based interventions similarly recalibrate bottom-up and top-down interoceptive processes, reducing anxiety and improving autonomic regulation (Sugawara et al., 2024; Quadt et al., 2021).

**Neuromodulators as gain controls**

Neuromodulatory systems regulate the precision-weighting of prediction errors, acting as gain controls that determine which sensory stream dominates integration. Within predictive coding, precision reflects the estimated reliability of a signal; neuromodulators adjust neuronal gain to amplify or attenuate prediction errors accordingly (Barrett & Simmons, 2015; Seth & Friston, 2016). This mechanism allows flexible reallocation of influence between interoceptive and exteroceptive inputs depending on task demands, context, and bodily state.

Experimental and computational work supports this role. Neuromodulators such as dopamine and norepinephrine dynamically adjust the excitability of cortical circuits, enhancing the salience of reliable signals while dampening noise. This allows descending predictions to engage autonomic or motor reflexes only when corresponding error signals have been precision-weighted, effectively filtering out unreliable inputs (Seth & Friston, 2016). Vagal stimulation provides direct evidence for neuromodulatory gain control. Both invasive VNS and non-invasive taVNS modulate interoceptive processing by strengthening vagal afferent pathways. TaVNS enhances heartbeat detection accuracy, increases heartbeat-evoked potentials (HEPs), and activates interoceptive hubs including the insula, ACC, and nucleus tractus solitarius (Villani et al., 2019; Müller et al., 2022; Richter et al., 2021). This supports its role in augmenting the precision of interoceptive signals and rebalancing body-brain integration.

Recent work also shows that interoceptive training enhanced by taVNS increases resting-state connectivity within bottom-up circuits (AIC-ACC-NTS) and top-down control pathways (AIC-DLPFC-SMG), while suppressing irrelevant coupling (AIC-visual cortex) (Sugawara et al., 2024). These findings suggest that neuromodulation strengthens both sensory precision and higher-order control, enabling adaptive recalibration of bodily awareness. Evidence also shows that taVNS enhances interoceptive processing at a representational (CNS) level rather than by altering basic autonomic variables. Studies confirm no significant changes in HR, HRV, or BP during stimulation (Villani et al., 2019; Richter et al., 2021), yet both accuracy and confidence in heartbeat detection improve, suggesting enhanced signal-to-noise ratios in interoceptive circuits (Villani et al., 2019; Richter et al., 2021). Beyond cardiac signals, taVNS strengthens stomach-brain coupling in the NTS and dopaminergic midbrain (VTA, SN), correlating with subjective hunger (Müller et al., 2022). This demonstrates vagal input as a discernible interoceptive signal, consistent with gut-brain reward pathways (Han et al., 2018; Fermin et al., 2023). Clinically, VNS has been shown to reduce pain perception in fibromyalgia, migraines, and depression, highlighting its role in interoceptive pain modulation (Lange et al., 2011; Barbanti et al., 2015; Borckardt et al., 2005; Paciorek & Skora, 2020). At the neuromodulatory level, distinct neurotransmitters implement gain functions that shape prediction error weighting. Dopamine encodes interoceptive precision and confidence (Fermin et al., 2023), norepinephrine regulates salience and attention,with VNS enhancing locus coeruleus activity (Follesa et al., 2007; Richter et al., 2021),and acetylcholine supports attentional control and flexible learning of interoceptive predictions (Fermin et al., 2023).

The interplay of these neuromodulatory systems is therefore central to successful interoceptive-exteroceptive integration and shapes the quality of the integrated percept (Barrett & Simmons, 2015). Dysregulation in neuromodulatory tone helps explain why identical bodily signals elicit divergent subjective experiences across individuals and clinical populations (Garfinkel et al., 2015; Khalsa et al., 2018; Owens et al., 2018; Paulus & Stein, 2010). Such dysfunctions are implicated in anxiety, depression, and post-traumatic stress disorder (Harshaw, 2015; Khalsa et al., 2018; Paciorek & Skora, 2020; Sugawara et al., 2024).

**Fear, Exteroception, and Healing via Interoceptive Rebalancing**

Fear is often initiated and maintained via exteroceptive stimuli, where external threat cues are afforded disproportionately high precision, leading to persistent hypervigilance. In PTSD, for example, exteroceptive triggers such as environmental sights and sounds become tightly bound to danger signals, producing exaggerated prediction error responses and heightened activation of sensory cortices and the salience network (anterior insula, ACC, amygdala) (Kearney & Lanius, 2022). The review on fear learning and PTSD further highlights that external cues act as the dominant conditioned stimuli in fear acquisition and generalization, reinforcing exteroceptive overweighting (Spalding, 2018; Lissek et al., 2014).

By contrast, interoceptive awareness has been identified as a critical resilience factor in trauma recovery. Somatic Experiencing (SE), a body-oriented therapy, explicitly works to restore balance by increasing interoceptive and proprioceptive attention through titration (introducing activation in small increments), pendulation (oscillating between safety and activation), and resourcing techniques. These interventions shift precision weighting back toward internal bodily cues, reducing exteroceptive dominance and mitigating hypervigilant states (Payne et al., 2015).

Fear reflects a model state in which exteroceptive prediction errors are overweighted relative to interoceptive priors, biasing arbitration toward the external stream. Interventions like SE and MBSR act by enhancing interoceptive precision and stabilizing priors, thereby rebalancing arbitration, improving integration, and supporting coherent appraisal and outcome generation.

**Updating, Action, and Appraisal: The Closed-Loop Mechanism**

Persistent prediction errors that survive precision-weighting drive updates to the brain's generative model, refining priors to better reflect bodily and environmental contingencies (Critchley & Garfinkel, 2017; Barrett & Simmons, 2015; Seth & Friston, 2016). Dopaminergic signaling encodes the confidence associated with these errors, facilitating the synaptic plasticity that underpins model revision (Fermin et al., 2023). Within the insular hierarchy, ascending prediction errors from granular and dysgranular modules are transformed into higher-order representations in the anterior insula and prefrontal cortices, where priors are stored and updated. The prefrontal cortex thus generates third-order interoceptive representations, integrating affective, contextual, and conceptual layers of experience (Fermin et al., 2023).

Prediction errors can also be minimized by acting on the body or environment to align sensory inputs with expectations. This principle of active inference is implemented via descending predictions engaging autonomic and motor systems (Friston, 2010). The anterior insula and ACC regulate breathing, heart rate, or posture to reduce mismatch between expected and actual states, while motor actions (e.g., approach or avoidance) modify exteroceptive context to restore coherence. These descending adjustments constitute the embodied component of predictive regulation, whereby cognition and physiology jointly resolve uncertainty.

Attention dynamically modulates which errors drive updates. The AIC-ACC salience network reallocates precision to the most contextually relevant stream, allowing flexible transitions between interoceptive and exteroceptive focus (Menon & Uddin, 2010; Seth, 2013). This flexibility maintains adaptive regulation: excessive interoceptive focus leads to anxious hypervigilance, whereas chronic neglect produces blunted affect and dysregulation (Nord & Garfinkel, 2022).

At the highest level, integrated interoceptive and exteroceptive representations are contextualized into subjective emotional appraisals by vmPFC and dorsomedial PFC. These regions generate a coherent bodily state within the body's environment (Critchley & Garfinkel, 2017; Seth & Friston, 2016). Successful alignment yields stability in self-perception and emotion; persistent misalignment generates conflicted appraisals manifesting as anxiety, dysphoria, or dissociation. Physiological indices such as HRV and vagal tone serve as observable markers of this coherence (Garfinkel et al., 2015; Greenwood & Garfinkel, 2025).

## Neural Markers of Interoceptive-Exteroceptive Coherence

### Reciprocal suppression between vmPFC and oPFC

Both fMRI and intracranial EEG reveal that interoceptive feedback preferentially activates vmPFC while suppressing oPFC, whereas exteroceptive input produces the opposite pattern (Iravani et al., 2024). This reciprocal competition indicates alternating dominance of internal versus external predictive streams, aligning with the model's arbitration principle.

### AIC as comparator and mismatch detector

False physiological feedback paradigms robustly increase right AIC activity, confirming its role in signaling interoceptive-exteroceptive mismatch (Gray et al., 2007; García-Cordero et al., 2017). Frequency dissociations such as high-frequency oscillations for interoception, low-frequency for exteroception support hierarchical segregation within the same circuitry.

### Amygdala as affective weighting hub

The vmPFC exerts inhibitory control over the amygdala; hypoactivation in vmPFC or OFC leads to disinhibited affective responses and fear overgeneralization (Spalding et al., 2018; Likhtik et al., 2014). Directional vmPFC-amygdala coupling is associated with successful discrimination of threat from safety, consistent with its role in adaptive emotional inference.

### Parallel but converging streams

Heartbeat-evoked potential (HEP) studies reveal distinct yet interacting interoceptive and exteroceptive effects converging in posterior cingulate cortex (Loescher et al., 2025), providing strong neural support for parallel hierarchical integration.

### Precision weighting and coherence metrics

High interoceptive sensitivity corresponds to stronger late ERPs (P300, slow wave) and increased AIC-ACC connectivity (Herbert et al., 2007). Meditation and biofeedback studies show enhanced network coherence, reduced prediction error noise, and improved emotion regulation (Jadhav et al., 2017; Tee & Leong, 2018). Conversely, maladaptive states such as PTSD, GAD, and schizophrenia display disrupted coherence and altered microstates (Lissek et al., 2014).

**Quantifiable coherence markers include:**

- Reciprocal vmPFC-oPFC suppression ratios (BOLD or power-based).
- AIC mismatch activity (amplitude or frequency shifts).
- vmPFC-amygdala theta-band synchrony as affective weighting index.
- Whole network coherence measures (magnitude-squared coherence, directed transfer function).

Balanced coherence reflects aligned integration (low noise, efficient updating, and emotional stability) whereas disrupted coherence marks maladaptive, high-noise states such as PTSD or depression.